\documentclass{emulateapj}

\usepackage{graphics}
\usepackage{epsfig}
\usepackage{natbib}
\shorttitle{COS Observations of 2M1207}
\shortauthors{France et al.}

\begin{document}


\title{Metal Depletion and Warm H$_{2}$ in the Brown Dwarf 2M1207 Accretion Disk\altaffilmark{*}}


\author{Kevin France\altaffilmark{1}} 
\affil{Center for Astrophysics and Space Astronomy, 389 UCB, University of Colorado, 
Boulder, CO 80309}
\author{Jeffrey L. Linsky}
\affil{JILA, University of Colorado and NIST, Boulder, CO 80309-0440}

\author{Alexander Brown}
\affil{Center for Astrophysics and Space Astronomy, 389 UCB, University of Colorado, 
Boulder, CO 80309}

\author{Cynthia S. Froning} 
\affil{Center for Astrophysics and Space Astronomy, Department of Astrophysical and Planetary Sciences389 UCB, University of Colorado, 
Boulder, CO 80309}

\author{and St\'{e}phane B\'{e}land}
\affil{Center for Astrophysics and Space Astronomy, 389 UCB, University of Colorado, 
Boulder, CO 80309}


\altaffiltext{*}{Based on observations made with the NASA/ESA $Hubble$~S$pace$~$Telescope$, obtained from the data archive at the Space Telescope Science Institute. STScI is operated by the Association of Universities for Research in Astronomy, Inc. under NASA contract NAS 5-26555.}

\altaffiltext{1}{kevin.france@colorado.edu}


\begin{abstract}
We present new far-ultraviolet observations of the young M8 brown dwarf 
2MASS J12073346-3932539, which is surrounded by an accretion disk.  The data were obtained using the $Hubble$~$Space$~$Telescope$-Cosmic Origins Spectrograph.  Moderate resolution spectra ($R$~$\approx$~17,000~--~18,000) obtained in the 
1150~--~1750~\AA\ and 2770~--~2830~\AA\ bandpasses reveal H$_{2}$ emission excited by \ion{H}{1} Ly$\alpha$ photons, several ionization states of carbon (\ion{C}{1}~--~\ion{C}{4}), and hot gas emission lines of \ion{He}{2} and \ion{N}{5} ($T$~$\approx$~10$^{4}$~--~10$^{5}$ K).
Emission from some species that would be found in a typical thermal plasma at this temperature (\ion{Si}{2}, \ion{Si}{3}, \ion{Si}{4}, and \ion{Mg}{2}) are not detected. The non-detections indicate that these refractory elements are depleted into grains, and that accretion shocks dominate the production of the hot gas observed on 2MASS J12073346-3932539.
We use the observed \ion{C}{4} luminosity to constrain the mass accretion rate in this system.
We use the kinematically broadened H$_{2}$ profile to confirm that the majority of the molecular emission arises in the disk, measure the radius of the inner hole of the disk ($R_{hole}$~$\approx$~3$R_{*}$), and constrain the physical conditions of the warm molecular phase of the disk ($T$(H$_{2}$)~$\approx$ 2500~--~4000~K).  A second, most likely unresolved H$_{2}$ component is identified.  This feature is either near the stellar surface in the region of the accretion shock or in a molecular outflow, although the possibility that this Jovian-like emission arises on the day-side disk of a 6~$M_{J}$ companion (2M1207b) cannot be conclusively ruled out.  In general, we find that this young brown dwarf disk system is a low-mass analog to classical T Tauri stars that are observed to produce H$_{2}$ emission from a warm layer in their disks, such as the well studied TW Hya and DF Tau systems.

\end{abstract}



\clearpage


\section{Introduction}

Brown dwarfs form the low-mass, low-temperature end of 
the galactic stellar mass distribution.  Optical and near-IR surveys carried 
out over the past two decades have detected hundreds 
of objects with masses below the sustained hydrogen burning threshold 
(about 0.07~M$_{\odot}$; Riaz \& Gizis 2007).~\nocite{riaz07}  
These brown dwarfs have effective temperatures
$T_{eff}$~$\lesssim$~2500 K in the spectral sequence from 
late M-stars through the L and T dwarfs.  
Although models of such low $T_{eff}$ brown dwarfs predict essentially 
no photospheric emission in the far-ultraviolet 
(far-UV; 1100~$\leq$~$\lambda$~$\leq$~1700~\AA), the far-UV spectra of
of these objects are mostly unexplored (see Hawley \& Johns-Krull 2003 and Gizis et al. 2005
for UV spectra of 4 late M-stars).  \nocite{hawley03,gizis05}

More massive M dwarfs (spectral types earlier than $\sim$~M5) have strong magnetic fields that, through processes still not fully understood, produce hot plasma and nonthermal particles. The rich
phenomenology includes chromospheres ($T$~$\sim$~10$^{4}$ K), transition regions ($T$~$\sim$~10$^{5}$ K), 
coronae ($T$~$\sim$~10$^{6}$ K), flares, and other transient activity. 
We observe this phenomenology through X-ray, far-UV, optical, and nonthermal radio emission lines
and continua. 
It is unclear if far-UV emission is common in brown dwarfs.  H$\alpha$ compared to 
L$_{bol}$ is observed to decrease in late M dwarfs (e.g., Berger et al. 2010), which would suggest less hot gas capable of producing far-UV emission.\nocite{berger10}
Alternatively, magnetically powered activity may extend through the late M dwarfs (and 
into the L and T classes), possibly with enhanced flaring relative to their more massive
counterparts~\citep{fleming00,berger08}.  Using $HST$-STIS, Hawley \& Johns-Krull (2003) found that M7-M9 stars 
show bright emission in C~IV $\lambda$1550 and other lines formed in high 
temperature plasmas. Also, Welsh et al. (2006) found both near-UV and far-UV 
emission and flares on several late-M dwarfs observed by $GALEX$.~\nocite{hawley03,welsh06} 

Far-UV emission from accretion shocks is another mechanism that could be important 
for young brown dwarfs where gas-rich accretion disks exist.  2MASS J12073346-3932539 (2M1207)
is an M8 brown dwarf ($M$~$\approx$~0.024~$M_{\odot}$; Riaz \& Gizis 2007) located in the $\sim$
10 Myr TW Hya association~\citep{kastner97,webb99}, at a distance of 52.4 pc (V = 20.15; Ducourant et al. 2008).\nocite{riaz07,ducourant08} Initial evidence for a circumstellar disk 
in this system was inferred from the association with the TW Hya group and the detection of strong H$\alpha$ emission, indicative of active accretion~\citep{gizis02}.  More recently, photometric and spectroscopic observations of 2M1207 from the far-red to the mid-infrared (IR) have allowed various groups to confirm the existence of a circumstellar dust disk~\citep{riaz07,morrow08}.  The accretion from this system has been shown to vary on timescales ranging from hours to weeks~\citep{scholz05,scholz06}, although the absolute level of the mass accretion rate ($\dot{M}$$_{acc}$) is somewhat unclear~\citep{herczeg09}.

While 2M1207 is an active accretor, searches for magnetic fields have returned only a surprisingly 
low 3-$\sigma$ upper limit.  Reiners et al. (2009) constrain the total magnetic flux, $Bf$~$<$~1 kG, where $B$ is the unsigned photospheric magnetic field, and $f$ is the magnetic field filling factor.~\nocite{reiners09}
This object has the highest quality existing far-UV dataset of any brown dwarf, obtained with $HST$-STIS~\citep{gizis05}, where several lines produced in hot gas (most notably \ion{He}{2} and \ion{C}{4}) were observed.  The observation of these lines, combined with a non-detection of \ion{Si}{4} emission, led these authors to conclude that the UV emission is produced by accretion, and that the silicon in the 2M1207 disk has depleted into dust grains.  If the hot gas lines were produced in 
a thermal plasma due to stellar activity, all astrophysically abundant species with similar emissivities should be present (for reasonable metallicities), the lack of \ion{Si}{4} implies that the hot gas is not primarily created in the stellar atmosphere, but at the shock interface between the accretion disk and the stellar surface.

\begin{figure*}
\begin{center}
\hspace{+0.0in}
\epsfig{figure=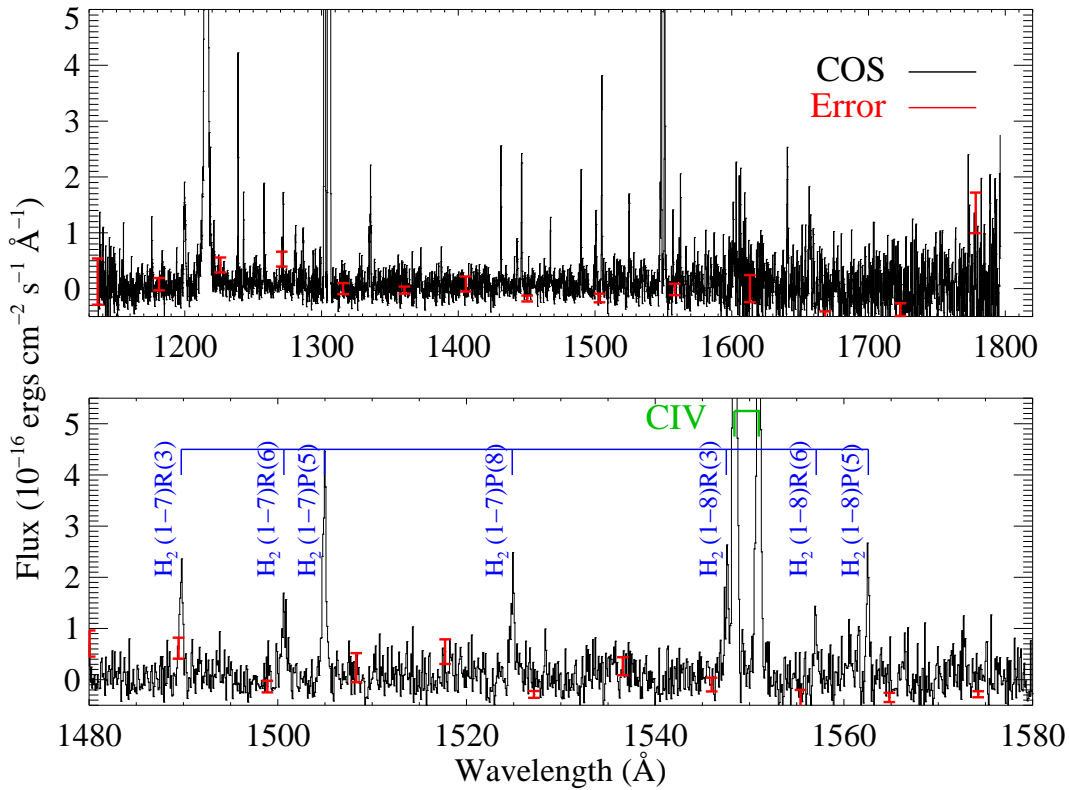,width=4.5in,angle=90}
\caption{\label{cosovly} The far-UV spectrum of 2M1207.  The top plot is a
weighted coaddition of the $HST$-COS G130M and G160M observations.  
The bottom plot highlights the 1480~--~1580~\AA\ region with relatively strong
lines of warm (H$_{2}$; $T$~$\gtrsim$~2500~K.) and hot (\ion{C}{4}; $T$~$\approx$~ 
1~$\times$~10$^{5}$ K) gas.
 }
\end{center}
\end{figure*}

\begin{deluxetable}{cccc}
\tabletypesize{\footnotesize}
\tablecaption{2M1207 COS observing log. \label{cos_obs}}
\tablewidth{0pt}
\tablehead{
\colhead{Dataset} & \colhead{COS Mode} & \colhead{Central Wavelength}   
& \colhead{T$_{exp}$ (s)} 
}
\startdata	
lb4p02fm	& 	G160M 	& 	1600		&	 1410	 \\
lb4p02ft	& 	G160M 	& 	1611 	&	 2975	 \\
lb4p02gn	& 	G160M 	& 	1623 	&	 2975	 \\
lb4p02gr	& 	G130M 	& 	1291 	&	 2967	 \\
lb4p02gx	& 	G130M 	& 	1300 	&	 3000	 \\
lb4p02h6	& 	G130M 	& 	1309 	&	 3005	 \\
lb4p01s1	& 	G285M 	& 	2676 	&	 3050  	 \\
lb4p01rx	& 	G140L 	& 	1230 	&	 3015  	 \\
lb4p01rv	& 	G140L 	& 	1230 	&	 3015  	 \\
lb4p01rt	& 	G140L 	& 	1230 	&	 200  	 \\
 \enddata

\end{deluxetable}

Molecular hydrogen (H$_{2}$) emission was also detected in the STIS observations, although the low spectral resolution of the G140L mode ($\Delta$$v$~$\sim$~240 km s$^{-1}$) prevented a conclusive determination of the location of the molecular gas in the system.  The most likely origin of the H$_{2}$ emission is in a warm molecular layer of the circumstellar disk, in analogy to the warm disks seen around more massive classical T Tauri stars (CTTSs; Herczeg et al. 2006).  It is worth noting in introducing the 2M1207 system that another reservoir of H$_{2}$ resides at the extrasolar giant planet companion, 2M1207b ($M_{b}$~$\sim$~6 $M_{J}$, secondary/primary mass ratio $\sim$~1:4; Chauvin et al. 2004; Song et al. 2006).~\nocite{chauvin04,song06}  This object is at a radial distance of $\sim$~40 AU, although the orbit is poorly constrained due to the long temporal baseline necessary for orbital monitoring.  The nature of 2M1207b is the subject of some debate~\citep{mohanty07,mamajek07,ducourant08}.



In this paper, we present new far-UV observations of this interesting low-mass system. 
These data cover a similar spectral bandpass as the $HST$-STIS observations presented by~\citet{gizis05}, but with order of magnitude increases in sensitivity and resolving power.
We describe the COS observations and data reduction in \S2.  A quantitative analysis of the far-UV spectrum is presented in \S3, with a focus on the properties of the H$_{2}$ in the circumstellar disk and the hot gas produced in the accretion shock.  In \S4, we put these results in the context of other young objects in a stage of active disk accretion and argue that the similarities are evidence that accreting brown dwarfs are low-mass analogs to CTTSs.  We conclude with a brief summary in \S5.

\section{$HST$-COS Observations and Data Reduction}

2M1207 was observed with the medium resolution far-UV modes (G130M and G160M) of $HST$-COS on 2009 December 08 for a total of 16333 seconds.  A description of the COS instrument and on-orbit 
performance characteristics are in preparation (J. C. Green et al. 2010, in preparation; S. N. Osterman et al. 2010, in preparation). ~\nocite{green10,osterman10}
In order to achieve continuous spectral coverage and minimize fixed pattern noise, observations in each grating were made at three central wavelength settings ($\lambda$1291,
1300, and 1309 in G130M and $\lambda$1600, 1611, and 1623 in G160M).  This combination of grating settings covers the 1140~$\leq$~$\lambda$~$\leq$~1790~\AA\ bandpass, at a resolving power of $R$~$\approx$~~17,000~--~18,000.
In addition, COS observed 2M1207 in the low-resolution G140L mode ($t_{exp}$ = 6230s) and the medium resolution near-UV mode G285M on 2009 December 03.  The $\lambda$2676 setting was used to provide coverage at the \ion{Mg}{2} 2800~\AA\ doublet, and the G285M exposure time was 3050 seconds.
Table 1 lists the COS data sets used in this work.

All observations were centered on 2M1207A (R.A. = 12$^{\mathrm h}$ 07$^{\mathrm m}$ 33.46$^{\mathrm s}$, Dec. = -39\arcdeg 32\arcmin 53.97\arcsec ; J2000) and COS performed a dispersed light target acquisition with the G160M grating.  The target coordinates and proper motions ($\mu$R.A. = -0.00555 s yr$^{-1}$, 
$\mu$Dec = 0.0226 \arcsec yr$^{-1}$; Epoch = 2006.7) were taken from~\citet{ducourant08}.
The data were originally processed with the COS calibration pipeline, CALCOS\footnote{We refer the reader to the cycle 18 COS Instrument Handbook for more details: {\tt http://www.stsci.edu/hst/cos/documents/handbooks/current/cos\_cover.html}} v2.11, and combined with a custom IDL coaddition procedure.  We found it necessary to reprocess the far-UV observations with a custom version of CALCOS because incomplete pulse-height screening produced residual spurious features in the coadded spectra.  
The COS G130M and G160M observations are presented in Figure 1.

\section{Analysis and Results} 

\subsection{Circumstellar Disk Profile from Warm H$_{2}$}

Initial far-UV observations~\citep{gizis05} did not have sufficient velocity resolution 
to set meaningful constraints on the kinematics of the 2M1207 system.  The COS observations analyzed here have a factor of $\approx$ 15 higher resolving power than those acquired with STIS G140L.
\citet{gizis05} noted the presence of H$_{2}$ lines pumped by \ion{H}{1} Ly$\alpha$ photons, for which they assumed a circumstellar origin.  We identify 14 clearly detected emission lines of H$_{2}$ (Table 2), all excited by Ly$\alpha$ through the coincident $B$~--~$X$ (1~--~2) R(6) 1215.73~\AA\ and (1~--~2) P(5) 1216.07~\AA\ absorbing transitions~\citep{shull78,valenti00}.  
These absorbing transitions are shifted from the Ly$\alpha$ rest frame by $\approx$ +15 and +100 km s$^{-1}$, respectively.  

The thermal width of a population of H$_{2}$ emitting gas will always be unresolved at the $\approx$ 15~--~20 km s$^{-1}$ velocity resolution of COS, hence evidence for a resolved velocity structure in the molecular lines can be attributed to kinematic broadening.  The velocity structure can in turn be interpreted as a physical structure for the emitting gas.  We found that while over a dozen H$_{2}$ lines were detected, the signal-to-noise (S/N) in a given emission line made line-profile fitting highly uncertain.  In order to improve the quality of the velocity fit, we created a normalized line profile from a coaddition of the six strongest H$_{2}$ lines [(1~--~3) R(3), (1~--~6) R(3), (1~--~6) P(5), (1~--~7) R(3), (1~--~7) P(5), and (1~--~8) P(5)].    The line profile was fit using a modified version of the IDL MPFIT function, customized to incorporate the COS linespread function (LSF\footnote{The COS LSF experiences a wavelength dependent non-Gaussianity due to the introduction of mid-frequency wave-front errors produced by the polishing errors on the $HST$ primary and secondary mirrors; {\tt http://www.stsci.edu/hst/cos/documents/isrs/}}).  
The model LSFs used here are based on numerical simulations of the $HST$ telescope and are 
qualitatively similar to a two Gaussian fit.  Our fitting routine convolves an underlying Gaussian profile with this LSF, and returns the amplitude, line center, and FWHM of the original Gaussian line shape (to measure the intrinsic H$_{2}$ line profile, the convolution with the Keplerian disk profile must also be taken into account).  However, for the broad lines observed in 2M1207, the COS LSF is expected to be virtually indistinguishable from Gaussian~\citep{parviz09}.  For the H$_{2}$ lines described here, we find a $\leq$~2\% relative difference between Gaussian and COS LSF fitting.

\begin{figure}
\begin{center}
\hspace{+0.0in}
\epsfig{figure=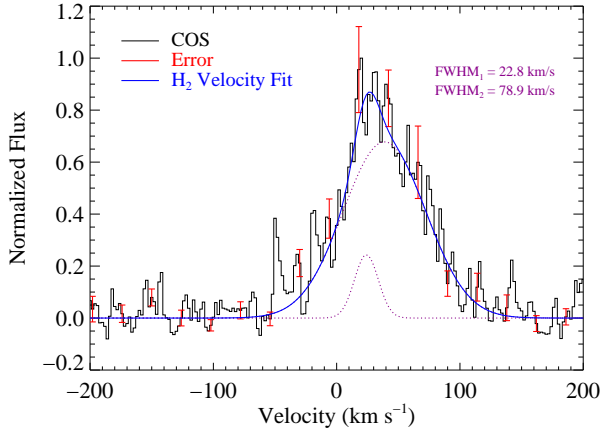,width=2.5in,angle=90}
\caption{\label{cosovly} The H$_{2}$ velocity profile created from a coaddition of the six strongest Ly$\alpha$ pumped lines in the far-UV spectrum of 2M1207.  A minimum of two components (fit with the appropriate COS linespread function) are needed to 
achieve a reasonable fit to the H$_{2}$ profile.  The broad component is representative of the kinematics of the circumstellar disk, indicating a pile-up of material at the inner wall of the disk at $\approx$ 3 $R_{*}$, corresponding to the disk sublimation radius.  The second component is most likely unresolved, and may be located in a molecular outflow or near the accretion hotspot on the stellar surface.  The solid line is the sum of the two components.
 }
\end{center}
\end{figure}

\begin{figure}
\begin{center}
\hspace{+0.0in}
\epsfig{figure=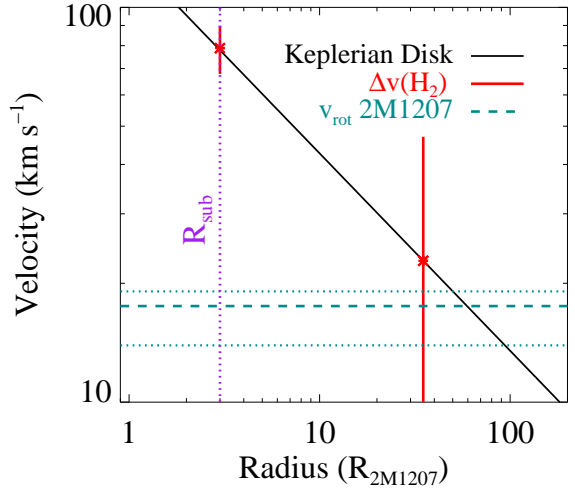,width=3.5in,angle=0}
\caption{\label{cosovly} H$_{2}$ velocity profiles in the context of the 2M1207 system.  The solid black curve is the rotational velocity of a Keplerian disk orbiting a 24 $M_{J}$ central object.  The two components (Figure 2) are shown in red with their corresponding error bars.  The broad component corresponds to the inner hole radius at the sublimation point.  The dashed and dotted dark cyan lines represent the nominal rotational velocity of 2M1207A and the rotational velocity limits assuming 1-$\sigma$ limits on $v$ sin $i$ and the inclination, respectively~\citep{reiners09,riaz07}.  This shows that the narrow H$_{2}$ component is consistent with emission from the surface of the primary, possibly at the accretion hotspot.
 }
\end{center}
\end{figure}

\begin{deluxetable}{lccc}
\tabletypesize{\footnotesize}
\tablecaption{Ly$\alpha$-pumped H$_{2}$ emission from the 2M1207 system. \label{bd_lines}}
\tablewidth{0pt}
\tablehead{
\colhead{Line ID\tablenotemark{a}} & \colhead{$\lambda_{rest}$} & \colhead{$\lambda_{obs}$\tablenotemark{b}
} & 
\colhead{Line Flux}    \\ 
    & (\AA) & (\AA) & (10$^{-16}$ ergs cm$^{-2}$ s$^{-1}$ )  }
\startdata
(1~--~2) P(8)	& 	1237.87 	& 1237.92 &	0.26 $\pm$ 0.11 		 \\
(1~--~3) R(3)	& 	1257.83 	& 1257.94 &	0.61 $\pm$ 0.12 		 \\
(1~--~3) R(6)	& 	1271.02 	& 1271.12 &	0.28 $\pm$ 0.11 		 \\
(1~--~3) P(5)	& 	1271.93 	& 1272.02 &	0.56 $\pm$ 0.07 		 \\
(1~--~6) R(3)	& 	1431.01 	& 1431.23 &	1.10 $\pm$ 0.23 		 \\
(1~--~6) R(6)	& 	1442.87 	& 1443.02 &	0.47 $\pm$ 0.14 		 \\
(1~--~6) P(5)	& 	1446.12 	& 1446.28 &	1.47 $\pm$ 0.37 		 \\
(1~--~6) P(8)	& 	1467.08 	& 1467.30 &	0.67 $\pm$ 0.60 		 \\
(1~--~7) R(3)	& 	1489.57 	& 1489.76 &	1.21 $\pm$ 0.12 		 \\
(1~--~7) R(6)	& 	1500.45 	& 1500.70 &	0.98 $\pm$ 0.14 		 \\
(1~--~7) P(5)	& 	1504.76 	& 1504.91 &	2.43 $\pm$ 0.15 		 \\
(1~--~7) P(8)	& 	1524.65 	& 1524.84 &	1.24 $\pm$ 0.18 		 \\
(1~--~8) R(3)	& 	1547.34 	& 1547.63 &	1.35 $\pm$ 0.26 		 \\
(1~--~8) P(5)	& 	1562.39 	& 1562.55 &	1.08 $\pm$ 0.25 		 \\
 \enddata


	\tablenotetext{a}{~Lyman band ($B$$^{1}\Sigma^{+}_{u}$~--~$X$$^{1}\Sigma^{+}_{g}$) transitions.} 
	\tablenotetext{b}{$\lambda_{obs}$ fits based on a flux weighted average of H$_{2}$ velocity component structure (Section 3.1). } 
\end{deluxetable}

In practice, the strongest H$_{2}$ lines fall near the center of the COS G160M segment B, so for the summed velocity profile, we employed the 1500~\AA\ LSF (we note that the COS LSF only changes slowly with wavelength).  It is immediately clear from a `by-eye' inspection of the resulting profile (Figure 2), that a single component is a poor representation of the H$_{2}$ velocity profile, and a $\chi^{2}$ analysis confirms this.  A two-component fit is the most conservative assumption as there is no additional evidence to imply a more complicated velocity structure.  Following the procedure described above, we found that the dominant velocity component was centered on $v_{1}$~=~+39 km s$^{-1}$ (relative to the rest wavelengths of H$_{2}$), with a velocity width (FWHM) $\Delta$$v_{1}$ = 79~$\pm$~11 km s$^{-1}$. 
A weaker, narrower component was identified at $v_{2}$~=~+24 km s$^{-1}$, with a velocity width $\Delta$$v_{2}$~= 23$^{+24}_{-23}$ km s$^{-1}$.  We discuss possible origins for this second feature in Section 4.2.  We note that while the relative velocities between different components should be robust, target acquisition errors cause zero point uncertainties (0~--~30 km s$^{-1}$) in the COS wavelength scale as applied by the current version of CALCOS.  However, none of the analysis presented here depends on the absolute velocity scale of the data.

We interpret the broad component as a tracer of the inner edge of the 2M1207 accretion disk~\citep{morrow08}.  The inner edge of the dust disk is set by the sublimation point of the grain population in the disk~\citep{whitney03a,whitney04}.  \citet{riaz07}, using broad band mid-infrared (IR) images from $Spitzer$ and models of low-mass disks, find that ISM-like grains with a maximum size of $a$~$\sim$~0.25 $\mu$m best approximate the observed properties of the 2M1207 disk (discussed further in Section 3.2.1).  Using these parameters, and sublimation temperature of 1600 K, they find a sublimation radius of $\sim$ 3 $R_{*}$.  If we assume a Keplerian disk profile, we find that our broad H$_{2}$ component is consistent with originating at this dust sublimation radius.  This is somewhat surprising because around more massive young stars with gas-rich disks, the inner gas radius is often observed to extend inward to the corotation radius, where the gas disk is 
truncated by stellar magnetic fields~\citep{najita07}.  2M1207 does not have a detectable magnetic field (Reiners et al. 2009; and see below for additional evidence against strong magnetic fields in this system), and it seems that the gas disk in 2M1207 is more closely tied to the dust disk relative to higher mass counterparts.  We display the relevant velocity widths and radii 
on a hypothetical disk velocity profile in Figure 3.

The excitation conditions required for Ly$\alpha$ excitation also constrains the molecular phase of the disk.  H$_{2}$ requires appreciable occupation in the $v$~=~2 level in order to absorb from the transitions coincident with Ly$\alpha$, while at lower temperatures different pumping transitions will dominate the resultant fluorescence spectrum.  At temperatures of a few hundred K, only the $v$~=~0 level is significantly populated, and one expects Ly$\beta$ (via $B$~--~$X$ (6~--~0) P(1)) to be the dominant excitation route.  At intermediate temperatures, $v$~=~1 is occupied, and \ion{O}{6} pumping becomes important (via $C$~--~$X$ (1~--~1) Q(3), see the Appendix), assuming the shock/magnetically energized radiation field is capable of producing this ion~\citep{france07b}.  Observations of Ly$\alpha$ pumped fluorescence indicate $T$(H$_{2}$)~$\gtrsim$~2500 K for the molecular component of the disk~\citep{herczeg04}.  The upper bound on the molecular disk temperature is set by the thermal dissociation threshold of H$_{2}$, $\approx$ 4000 K~\citep{shull82}.  Thus, our observations imply that the 2M1207 disk has a warm molecular component with a temperature in the range 2500~--~4000 K.

\subsection{Hot Gas (\ion{C}{4} and \ion{N}{5}) Velocity Profiles}

Emission from \ion{C}{4} $\lambda\lambda$ 1548 and 1550~\AA\ is the strongest feature in the far-UV spectrum
of 2M1207 that is not contaminated by telluric airglow.  The \ion{C}{4} 1548 profile suggests a double peak structure, however the S/N is not high enough to make a conclusive determination within the error bars.  
In order to obtain a more robust velocity profile of the hot gas ($T$~$\sim$~10$^{5}$ K) in 2M1207, 
we follow the procedure outlined above for the H$_{2}$ lines and coadd the lines of the \ion{C}{4} and \ion{N}{5} doublets.  Figure 4 shows this coadded profile, including a fit to the data.  We see that the suspected line structure is real, however the S/N still prevents an unambiguous interpretation.  We carried out a $\chi^{2}$-minimization analysis to determine the most likely underlying line profile.  We found that a two-component fit to the hot gas profile (Figure 4) produced a better fit (reduced $\chi^{2}$, $\chi_{red}^{2}$ = 1.828) than either a single emission component ($\chi_{red}^{2}$ = 1.992) or a single emission line with a superimposed absorption component ($\chi_{red}^{2}$ = 2.370).  While it may be possible to improve the fits by including additional component structure, the data do not support a more complicated interpretation.
We find that the hot gas has velocity components $v_{hot 1,2}$ = +22, 41 km s$^{-1}$, with $\Delta$$v_{hot 1,2}$~=~36, 76 km s$^{-1}$.  The broad component is most likely tracing material infalling along the accretion stream.  It is interesting to note that the hot gas profile is qualitatively similar to the time-variable H$\alpha$ profiles presented by~\citet{scholz05}.
The H$\alpha$ profiles show a broad emission line with a narrower, redshifted absorption component superimposed.  This line profile may be present for the far-UV hot gas lines, but considerably higher S/N is needed to test this possibility.

\begin{figure}
\begin{center}
\hspace{+0.0in}
\epsfig{figure=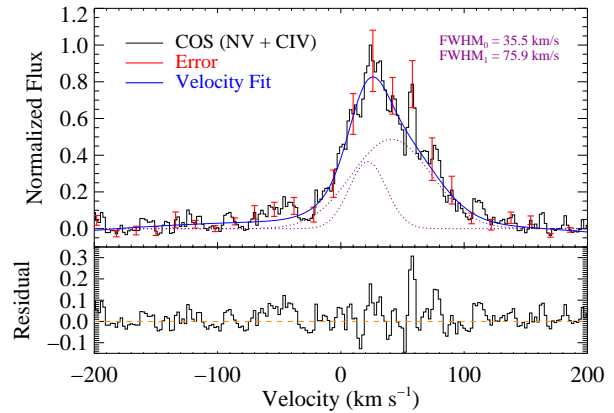,width=2.5in,angle=90}
\caption{\label{cosovly} The hot gas velocity profile of the accretion shock in 2M1207.
To improve the S/N in the profile, both lines of the \ion{C}{4} and \ion{N}{5} doublets were coadded, 
and the components fit using the COS linespread function. 
This fit has a reduced $\chi^{2}$ = 1.828, which was a better fit than either a single emission component or an emission component with a narrow red-shifted absorption component superimposed.
The solid line is the sum of the two components.
The lower panel displays the residuals of the fit.
 }
\end{center}
\end{figure}

\begin{figure}
\begin{center}
\hspace{+0.0in}
\epsfig{figure=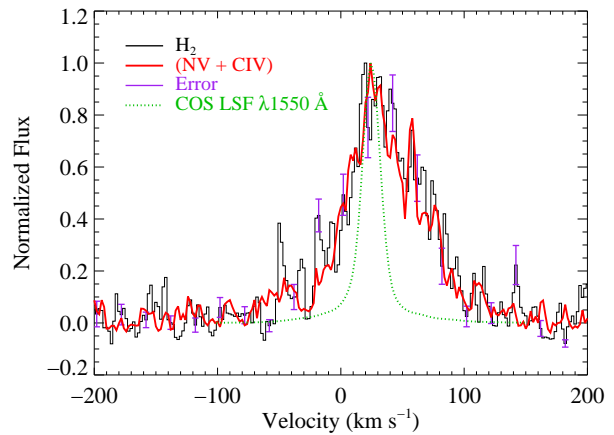,width=2.5in,angle=90}
\caption{\label{cosovly} A comparison of the warm (H$_{2}$) and hot (\ion{C}{4} + \ion{N}{5}) 
gas profiles presented in Figures 2 and 4.  The COS line spread function (offset by +24.1 km s$^{-1}$;  \S3.1) is shown as the dotted green 
line.  While the warm and hot gas profiles are qualitatively similar, they are most likely created by different physical processes (rotation vs. accretion), and are not related to the instrumental profile.
 }
\end{center}
\end{figure}

Figure 5 shows a comparison of the warm (H$_{2}$) and hot (\ion{C}{4} + \ion{N}{5}) gas profiles described above.  The COS LSF computed for $\lambda$~=~1550~\AA\ ($\lambda_{rest}$ \ion{C}{4}) is shown overplotted in green.  This clearly shows that the line profiles are fully resolved, and the LSF does not significantly alter the observed profiles, as expected for most emission lines observed with COS (S. N. Osterman et al. 2010, in preparation).  While it is interesting that the H$_{2}$ and hot gas profiles are qualitatively similar, we consider it most likely that this is coincidental and that the profiles are governed by different physical processes.  While we favor the interpretation that the H$_{2}$ is tracing the Keplerian rotation of the inner disk hole, it should be noted that the red wing of the H$_{2}$ profile could include a contribution 
from the infalling accretion stream.  This component would necessarily be in the outer regions of the accretion stream as H$_{2}$ will be collisionally dissociated in a \ion{C}{4} emitting plasma.  

In addition to \ion{C}{4} and \ion{N}{5}, we observe a range of ionization states of carbon, including its neutral form, and \ion{He}{2}.   These lines were fit using the emission profile observed in the summed \ion{N}{5} and \ion{C}{4} profiles as a proxy, and line strengths are presented in Table 3.  The spectra also include emission from Ly$\alpha$ and the \ion{O}{1} $\lambda$1304 multiplet, however the large aperture of COS does not permit us to separate the brown dwarf signal from geocoronal airglow emission.

\subsubsection{Limits on Dust Depleted Species: Si and Mg}


In a plasma with the range of temperatures necessary to excite \ion{C}{1} as well as the three ionization states of carbon described above, one would expect strong line emission from other astrophysically abundant species with similar excitation energies, specifically silicon and magnesium.  The COS observations presented here include wavelengths with strong emission lines of \ion{Si}{3} ($\lambda$1206~\AA), \ion{Si}{4} ($\lambda\lambda$ 1394 and 1403~\AA), and \ion{Mg}{2} ($\lambda\lambda$ 2796 and 2803~\AA).  We do not detect any of these species (Figures 6 and 7), and the very low detector background of the COS MCP allows us to put tight limits on the flux in the silicon lines.  Assuming the velocity width of the combined hot gas profile presented above, we can place 1-$\sigma$ integrated line flux limits of 
[0.18, 0.17, 0.19, 0.34, 0.29, 6.94, and 6.94] $\times$ 10$^{-16}$ ergs cm$^{-2}$ s$^{-1}$ for the [\ion{Si}{3} 1206, \ion{Si}{4} 1394, \ion{Si}{4} 1403, \ion{Si}{2} 1526, \ion{Si}{2} 1533, \ion{Mg}{2} 2796, and \ion{Mg}{2} 2803] transitions, respectively (Table 3).  
Figure 6 shows the expected emission profiles from \ion{Si}{3} and \ion{Si}{4} based on the observed flux of the \ion{C}{4} emission and the relative emissivities of the relevant transitions (assuming collisional ionization and solar abundances; Dere et al. 1997, 2009).~\nocite{dere97,dere09}
While the non-detection of silicon is highly significant, a direct comparison cannot be made between the observed \ion{C}{4} flux and the expected level of \ion{Mg}{2} emission as the species traditionally trace different atmospheric regions (chromosphere vs. transition region) in low mass stars.  We note however that observations of more massive M dwarfs find \ion{Mg}{2} emission to be much stronger ($\times$ $\sim$ 10) than that of \ion{C}{4}~\citep{byrne89}.
The absence of \ion{Si}{4} emission was observed by~\citet{gizis05} in the STIS observations of 2M1207, and COS allow us to set upper limits that are smaller by approximately an order of magnitude. 
We measure a \ion{C}{4}/\ion{Si}{4} ratio $\geq$ 35, very similar to the high ratios observed in some CTTSs, but significantly different from the values of $\sim$ unity that are observed in higher mass cool star atmospheres (Ayres et al. 1997; and see Herczeg et al. 2002, Section 4 for a discussion).~\nocite{ayres97,herczeg02}

\begin{figure}
\begin{center}
\hspace{+0.0in}
\epsfig{figure=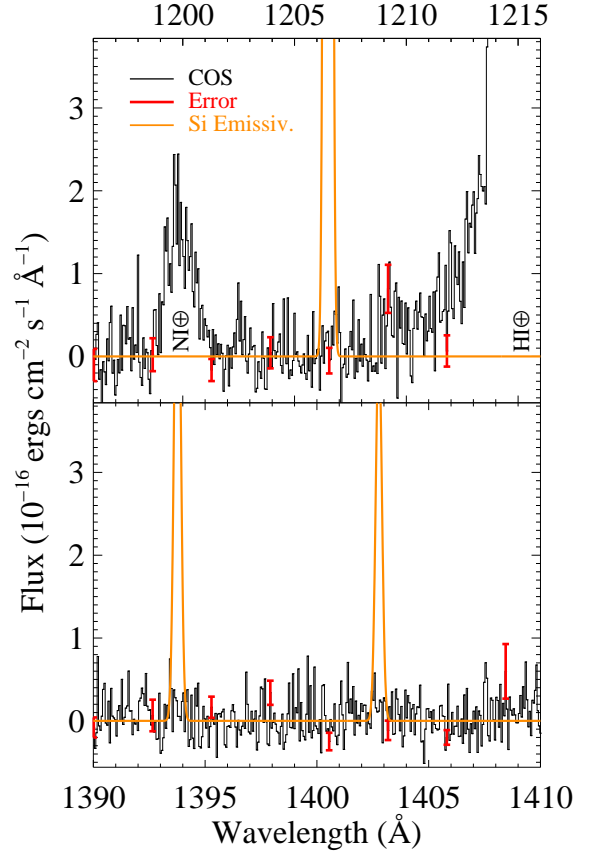,width=3.5in,angle=0}
\caption{\label{cosovly} G130M and G130M + G160M spectral regions where prominent emission lines
of \ion{Si}{3} ($\lambda$1206~\AA) and \ion{Si}{4} ($\lambda\lambda$1394 and 1403~\AA) would be expected if the hot gas in the 2M1207 system is created by a magnetically active upper atmosphere.  The orange spectrum is the expected Si profile based on the ratio of Si/\ion{C}{4} emissivities~\citep{dere09} for solar abundances at the observed \ion{C}{4} emission level (Table 3).
These non-detections imply that the silicon in the 2M1207 accretion disk has depleted into grains, and that accretion shocks produce the \ion{C}{4} and \ion{N}{5} observed in the system.
 }
\end{center}
\end{figure}

\begin{figure}
\begin{center}
\hspace{+0.0in}
\epsfig{figure=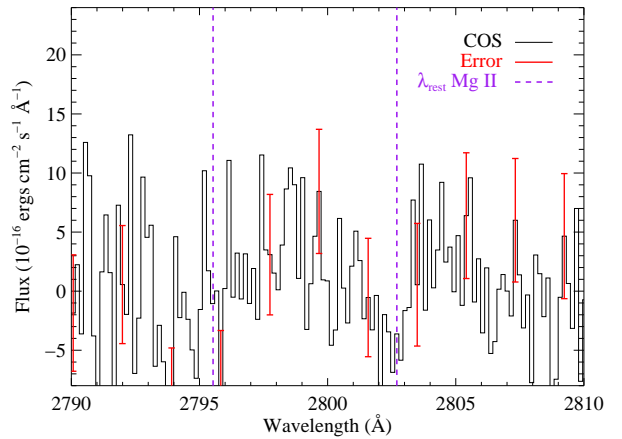,width=2.5in,angle=90}
\caption{\label{cosovly} Same as Figure 6, but for the \ion{Mg}{2} region observed with $HST$-COS G285M.
Upper limits on the line strengths of the \ion{Mg}{2} doublet are presented in Table 3.
 }
\end{center}
\end{figure}

Silicon and magnesium can be heavily depleted into dust grains~\citep{savage96}, and young circumstellar disks are known to exhibit grain growth~\citep{apai05} that can be a reservoir for refractory elements originally in the gas phase.  Evidence for grain growth can be seen in mid-IR spectra of disks where the 10 $\mu$m silicate emission feature has broadened or disappeared.  The mid-IR dust spectrum is somewhat ambiguous for 2M1207.  \citet{riaz07} claim a detection of this feature, using ground-based and $Spitzer$ (IRAC and MIPS) photometry to infer excess emission at 10 $\mu$m based on the ratio of 8.7/10.4 $\mu$m flux.  However, \citet{morrow08} used direct spectroscopic observations with the $Spitzer$-IRS to rule out any emission from the 10 and 20 $\mu$m silicate features.  The latter observation implies that the grains in the 2M1207 circumstellar disk have experienced significant evolution towards larger grains ($a$~$>$ 5 $\mu$m) and have most likely settled into the disk midplane~\citep{dullemond04}.  
This scenario is consistent with our non-detection of dust depletion species in the COS spectra of 2M1207, 
and is not particularly surprising given the $\sim$~10 Myr age of the TW Hya association~\citep{gizis02}.  \citet{sargent09}  discuss a survey of 65 TTSs in the Taurus-Auriga star forming region.  These disks show evidence of grain growth in a population of more massive disks that are appreciably younger than those in TW Hya.
In any event, this result presents an interesting constraint on the gas and dust composition in the disk, and provides additional evidence that accretion (and not magnetic activity) produces the hot gas in the 2M1207 system.  Figure 8 presents a cartoon representation of the inner region of the system based on the spectroscopic analysis presented in \S3.1 and \S3.2, as well as existing interpretation from the literature cited above.

\subsection{Emission Line Variability}

\citet{scholz05} and \citet{scholz06} report on the variability of H$\alpha$ emission from the 2M1207 system on timescales from hours to weeks, most likely due to variability in the accretion rate onto the stellar surface.  The COS far-UV MCP is a photon-counting detector, and data are recorded in ``time-tagged'' mode: an [$x,y,time$] coordinate is recorded for each observed photon.  This means that time variability in all spectral features can be tracked over the course of the observation, by isolating the appropriate $x,y$ coordinates in a coadded two-dimensional spectral image and summing the photons in that region over a given time step.  
We isolated three spectral regions to explore time variability in the 2M1207 data: 
\ion{N}{5} (COS mode: G130M, segment B), \ion{C}{4} (COS mode: G160M, segment B), and 
H$_{2}$ (1425 $\leq$ $\lambda$ $\leq$ 1530~\AA, COS mode: G160M, segment B).  We note that the (1~--~8) R(3) line of H$_{2}$ ($\lambda$ 1547.34~\AA) is included in the \ion{C}{4} region.  

\begin{deluxetable}{lccc}
\tabletypesize{\footnotesize}
\tablecaption{Atomic emission from the 2M1207 accretion shock. \label{bd_lines}}
\tablewidth{0pt}
\tablehead{
\colhead{Line ID} & \colhead{$\lambda_{rest}$} & \colhead{$\lambda_{obs}$\tablenotemark{a}
} & 
\colhead{Line Flux}    \\ 
    & (\AA) & (\AA) & (10$^{-16}$ ergs cm$^{-2}$ s$^{-1}$ )  }
\startdata
\ion{O}{6}\tablenotemark{b}	& 	1031.91 	& $\cdots$ &	$\leq$ 5.1		 \\
\ion{O}{6}	& 	1037.61 	& $\cdots$ &	$\leq$ 5.1		 \\
\ion{C}{3}	& 	1176 	& 1175.86 &	0.70 $\pm$ 0.34 		 \\
\ion{Si}{3}  & 	1206.50 	& $\cdots$ &	$\leq$~0.18		 \\
\ion{H}{1}$\oplus$\tablenotemark{c,}\tablenotemark{d}	& 	1215.67 	& 1215.66 &	2.3~$\times$~10$^{4}$ 		 \\
\ion{N}{5}  & 	1238.82 	& 1238.94 &	1.28 $\pm$ 0.17 		 \\
\ion{N}{5}  & 	1242.80 	& 1242.92 &	0.59 $\pm$ 0.16 		 \\
\ion{C}{1}?  & 1280.33 	& 1280.92 &	0.61 $\pm$ 0.12 		 \\
\ion{O}{1}$\oplus$\tablenotemark{d}  & 	1302.17 	& 1302.21 &	79.88 $\pm$ 1.59 		 \\
\ion{O}{1}$\oplus$  & 	1304.86 	& 1304.97 &	49.90 $\pm$ 3.96 		 \\
\ion{O}{1}$\oplus$  & 	1306.03 	& 1306.05 &	19.63 $\pm$ 1.75 		 \\
\ion{C}{2}\tablenotemark{d}  & 	1334.53 	& 1334.63 &	0.73 $\pm$ 0.40 		 \\
\ion{C}{2}  & 	1335.71 	& 1335.89 &	0.86 $\pm$ 0.09 		 \\
\ion{Si}{4}  & 	1393.76 	& $\cdots$ &	$\leq$~0.17		 \\
\ion{Si}{4}  & 	1402.77 	& $\cdots$ &	$\leq$~0.19		 \\
\ion{Si}{2}  & 	1526.71 	& $\cdots$ &	$\leq$~0.34		 \\
\ion{Si}{2}  & 	1533.43 	& $\cdots$ &	$\leq$~0.29		 \\
\ion{C}{4}  & 	1548.19 	& 1548.38 &	8.22 $\pm$~0.37	 \\
\ion{C}{4}  & 	1550.77 	& 1550.95 &	5.41 $\pm$~0.21		 \\
\ion{He}{2}  & 	1640.40 	& 1640.59 &	2.28 $\pm$ 0.54		 \\
\ion{C}{1}  & 	1657 	& 1657.69 &	4.83 $\pm$ 2.34		 \\
\ion{Mg}{2}  & 	2795.73 	& $\cdots$ &	$\leq$~6.94		 \\
\ion{Mg}{2}  & 	2802.70 	& $\cdots$ &	$\leq$~6.94		 \\
 \enddata


	\tablenotetext{a}{$\lambda_{obs}$ fits based on a flux weighted average of hot gas velocity component structure (Section 3.2). } 
	\tablenotetext{b}{An alternative upper limit on \ion{O}{6} $\lambda$ 1032~\AA\ derived from the absence of \ion{O}{6} pumped H$_{2}$ emission in the 2M1207 spectra is approximately 9.5~$\times$~10$^{-16}$ ergs cm$^{-2}$ s$^{-1}$ (see the Appendix for details). } 
\tablenotetext{c}{ 
~Lines labeled $\oplus$ are contaminated by geocoronal emission. } 
\tablenotetext{d}{Emission line flux decreased by interstellar absorption.} 

\end{deluxetable}

We do not find significant variations in the features tracked over the course of the observations.  
Figure 9 displays the emission line strengths as a function of exposure time.  We also plot a time spectrum of the background level, obtained over an extraction box with the same dimensions as that used for H$_{2}$, but offset by -50 pixels in the cross-dispersion direction.  
The time sampling was chosen to be 200 s, as this was the smallest interval that provided at least one background count in each time step.  Figure 9 shows what might be variability in the relatively weak \ion{N}{5} and H$_{2}$ lines, but a close inspection shows this change to simply be background variations, presumably related to the orbital position of $HST$.  The background flux level is $\lesssim$ 10$^{-4}$ counts s$^{-1}$ pixel$^{-1}$, similar to the background level of the B-segment COS MCP reported by~\citet{mccandliss09}.

For simplicity, we plot the \ion{N}{5}, \ion{C}{4}, and H$_{2}$ on the same time axis, but in practice the G130M observations were acquired on the $HST$ orbits following the G160M observations.  The \ion{N}{5} time sequence started roughly 3.6 hr ($\sim$ 0.15 $P_{rot}$) after the \ion{C}{4} and H$_{2}$ data.  If significant changes were seen in any of the lines, this might warrant further attention, but as all of the features were approximately constant, this treatment conveys the relevant information.  
H$_{2}$ is a proxy for the strength of the Ly$\alpha$ radiation field, and the constancy of the H$_{2}$ lines suggests that Ly$\alpha$ was roughly constant over the $\sim$~2.1 hr of the G160M observations.  One might expect Ly$\alpha$ to directly trace the other hot gas lines created in the accretion shock, and Figure 10 shows a comparison of the Ly$\alpha$ pumped H$_{2}$ emission and that of the highest S/N hot metal emission, \ion{C}{4}.  A correlation analysis finds a Pearson coefficient of 0.15, showing that correlated changes in the \ion{C}{4} and Ly$\alpha$ were not present during the G160M observations.

\section{Discussion}

\begin{figure*}
\begin{center}
\hspace{+0.0in}
\epsfig{figure=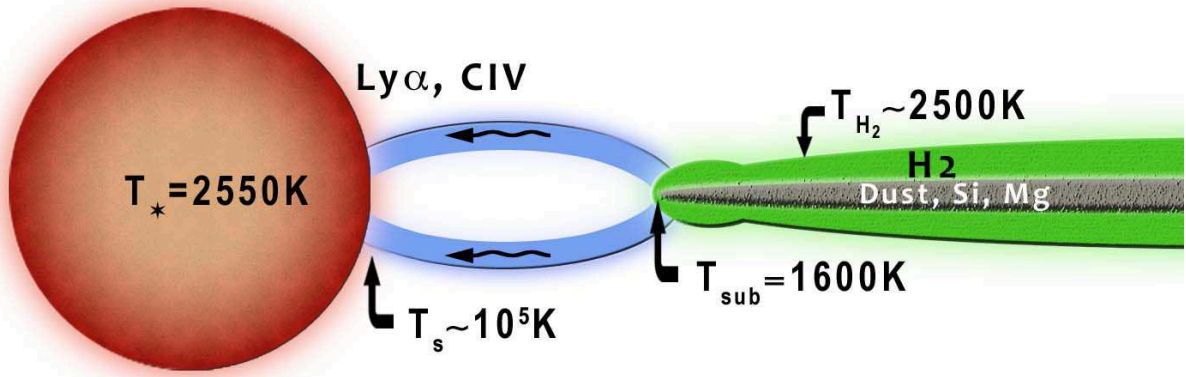,width=6.5in,angle=0}
\caption{\label{cosovly} A cartoon representation of the inner disk region of 2M1207, 
based on interpretations in the literature and the new $HST$-COS observations presented here.
The distances are not to scale.  In this cartoon, we see hot gas emission (Ly$\alpha$ and \ion{C}{4}, $T_{s}$~$\sim$~10$^{5}$K) produced where the accretion stream makes contact with the brown dwarf surface, dust and depleted species (Si and Mg) distributed near the mid-plane, while the Ly$\alpha$-pumped H$_{2}$ traces a warmer ($T(H_{2})$~$\approx$~2500~--~4000 K), extended surface layer of the disk.  The dust sublimation point is assumed to be $T_{sub}$~=~1600 K~\citep{riaz07}.
 }
\end{center}
\end{figure*}

\subsection{Mass Accretion Rate: \ion{C}{4} Luminosity }

We do not have contemporaneous observations of 2M1207 in the optical or NUV, where traditional accretion diagnostics are located (Herczeg et al. 2009 and references therein).~\nocite{herczeg09} \citet{krull00} present empirical relations for determining the $\dot{M}$$_{acc}$ of CTTSs, and while these relations were created for higher mass objects with larger mass accretion rates, it is interesting to compare an extrapolation of the CTTS relation to a more standard technique.  For this purpose, we use Equation 2 of~\citet{krull00}.
In addition to 2M1207 having over an order of magnitude smaller mass than any stars considered by~\citet{krull00}, we note they assume that the accretion emission is in excess of a saturated magnetic component that produces a surface \ion{C}{4} flux level of $F_{CIV}$ $>$ 10$^{6}$ ergs cm$^{-2}$ s$^{-1}$.  Taking a distance of 52.4 pc~\citep{ducourant08} and a stellar radius of 0.24~$R_{\odot}$, we find a total \ion{C}{4} surface flux of $F_{CIV}$ = 1.28 $\times$ 10$^{5}$ ergs cm$^{-2}$ s$^{-1}$ (where no saturated magnetic component has been subtracted).
While this is lower than the saturation threshold suggested by~\citet{krull00}, the combination of the low magnetic field at 2M1207 ($<$ 1 kG; Reiners et al. 2009) and the non-detection of Si and Mg species (Section 3.2.1) lead us to assert that essentially all of the \ion{C}{4} emission from 2M1207 is produced by accretion.

The empirical relation between the \ion{C}{4} luminosity ($L_{CIV}$, in units of ergs s$^{-1}$) and $\dot{M}$$_{acc}$ depends strongly on the method and values used for dereddening the observed fluxes, particularly at the wavelength of \ion{C}{4} (1550 \AA), where the effects of interstellar extinction are large~\citep{ccm}.  
The mass accretion rate is then 
\begin{equation}
log_{10}(\dot{M}_{acc}) = 0.753 \ log_{10}(L_{CIV}) - 29.89
\end{equation}
The $L_{CIV}$ is calculated to be 4.49 $\times$ 10$^{26}$ ergs s$^{-1}$
We note that the 2M1207 sightline is generally assumed to suffer no interstellar extinction ($A_{V}$~=~0.0; Herczeg et al 2009), and no correction was applied to the \ion{C}{4} line fluxes presented in Table 3.  We find log$_{10}$ $\dot{M}$$_{acc}$~$\approx$~-9.8 [$\dot{M}$$_{acc}$ = 1.6~$\times$~10$^{-10}$ $M_{\odot}$ yr$^{-1}$] from the \ion{C}{4} observations of the 2M1207 system.  This value is 
is consistent with the accretion level of 2M1207 derived from H$\alpha$ observations (log$_{10}$ $\dot{M}$$_{acc}$~=~-10.1~$\pm$~0.7) obtained in the ``high'' state~\citep{scholz05}.  \citet{krull00} note that alternative calibrations produce accretion rates that about 10 times lower than those given in Equation 1 above.  If that scaling is applied, we find that the $\dot{M}$$_{acc}$
derived from the \ion{C}{4} line strengths is consistent with the lower values observed by Scholz et al. (2005; log$_{10}$ $\dot{M}$$_{acc}$ = -10.8~$\pm$~0.5).

Interestingly, while we find the \ion{C}{4}-based accretion rate to be in excellent agreement with that derived from H$\alpha$ observations, our low value are approximately an order of magnitude greater than those measured using deep, low-resolution observations of the Balmer continuum (log$_{10}$ $\dot{M}$$_{acc}$~=~-11.9; Herczeg et al. 2009).  The accretion rate in the 2M1207 system is known to vary by at least an order of magnitude, and since none of observations were acquired simultaneously, it is plausible that variability causes the discrepancy between the mass accretion rates measure by H$\alpha$, \ion{C}{4}, and Balmer continuum observations.  Alternatively, 
absorption of Balmer continuum emission by the edge-on disk may lead to a lower estimation of the accretion rate by this method.


\subsection{Physical Origin of the Narrow H$_{2}$ Component}

In Figure 2, we displayed the coadded H$_{2}$ emission line profile of the six lines with the highest 
S/N in the COS M-grating data.  In Section 3.1, we discussed the dominant broad component and identify it as emission from a pile-up of material at the inner wall of the circumstellar disk, approximately at the disk sublimation radius (Figure 8).  The velocity width of the second component is poorly constrained as the fit is dominated by the broader, stronger component.  The velocity width is consistent with being an unresolved feature.  There are several possible physical origins for an unresolved H$_{2}$ population.  The most likely scenario seems to be that this additional emission arises at the stellar surface.  The photospheric temperature ($T_{*}$) is 2550 K~\citep{riaz07}, ideal for maintaining an H$_{2}$ population that is capable of being pumped by Ly$\alpha$ photons in thermal equilibrium.  If a photospheric origin is the correct interpretation, this would argue that the emitting molecules are near the accretion hotspot created at the interface of the infalling material from the disk, seen in our COS observations through several ionization states of He, C, and N.  The relaxation time for the electronic transitions of H$_{2}$ is very short ($A_{TOT}$ for the (1~--~2) R(6) transition coincident with Ly$\alpha$ is 1.68~$\times$~10$^{9}$ s$^{-1}$; Abgrall et al. 1993), and the UV transitions of the H$_{2}$ molecules would not be visible if they were not being actively excited.~\nocite{abgrall93a}  

The $\approx$ 15 km s$^{-1}$ blueshift of this component relative to the bulk of the H$_{2}$ emission from the disk suggests a possible outflow origin.  The CTTSs T Tau and RU Lupi show narrow, blueshifted H$_{2}$ emission that is thought to be indicative of a bipolar outflow~\citep{herczeg06}.  The blueshift of the outflow emission in these objects is roughly the same ($v$~=~-12 km s$^{-1}$) as that found for 2M1207.
If an outflow is the correct interpretation, the 15 km s$^{-1}$ relative velocity of the narrow H$_{2}$ component in 2M1207 is surprising because T Tau and RU Lupi host nearly face-on disks, where the outflow jet is pointed more directly at the observer.  It seems unlikely that the edge-on orientation of the 2M1207 disk would permit the same magnitude of blueshift produced in more massive, face-on disks, however, 2M1207 is observed to have [\ion{O}{1}] emission that is consistent with an outflow~\citep{whelan07}.  One final possibility is that the weak H$_{2}$ emission originates in the dayglow or aurorae of the 6~$M_{J}$ companion, 2M1207b (Chauvin et al. 2004; and see France et al. 2010 for a detailed discussion of the predicted UV emission properties of extrasolar giant planets).~\nocite{chauvin04,france10a}  The far-UV spectrum of Jupiter is dominated by H$_{2}$ emission, where the excitation is caused by electron-impact where the magnetic field lines connect to the planetary surface near the poles and solar-induced Ly$\beta$ fluorescence in the equatorial regions~\citep{feldman93,wolven98}.  In the instance of an additional energy source (in this case the Shoemaker Levy 9 impact), the Jovian atmosphere supports Ly$\alpha$ pumped H$_{2}$ emission~\citep{wolven97} similar to that observed in 2M1207.  The velocity shift due to the orbital motion of the planet would be undetectable at the COS resolution ($v_{orb}$~$\sim$~0.7 km s$^{-1}$ at 40 AU, assuming a circular orbit), and this scenario would require both a mechanism to heat the 2M1207b atmosphere to $T$~$\gtrsim$~2500 K, and produce a 15 km s$^{-1}$ outflow.  While we favor a photospheric origin for the narrow H$_{2}$ component in 2M1207, we cannot conclusively rule out an outflow or the extrasolar giant planet companion
as possible sources of the observed H$_{2}$.

\begin{figure}
\begin{center}
\hspace{+0.0in}
\epsfig{figure=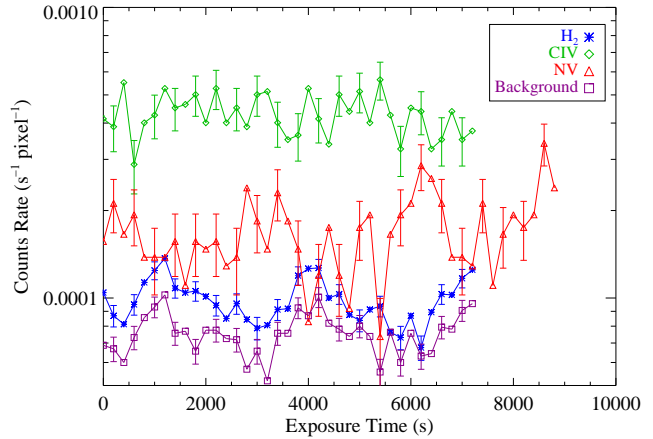,width=2.5in,angle=90}
\caption{\label{cosovly} Time-tagged fluxes from emission lines tracing the 
warm (H$_{2}$) and hot (\ion{C}{4} and \ion{N}{5}) components of the 2M1207 system, in 200 second time intervals.
The line fluxes are essentially constant, with most of the variability in H$_{2}$ and 
\ion{N}{5} caused by a time variable background level.  
 }
\end{center}
\end{figure}

\begin{figure}
\begin{center}
\hspace{+0.0in}
\epsfig{figure=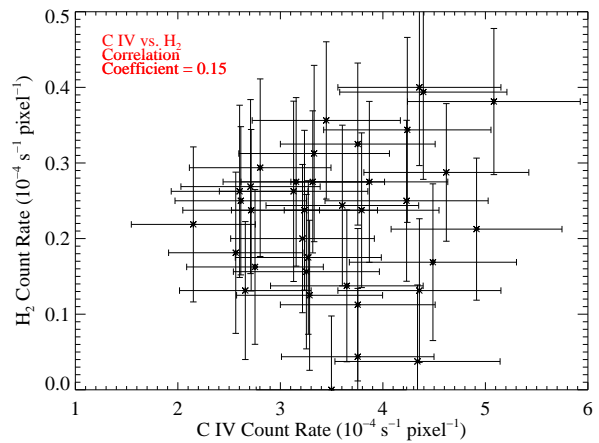,width=2.5in,angle=90}
\caption{\label{cosovly} A direct comparison of the background subtracted line fluxes ($\Delta$$t$~=~200 s) of \ion{C}{4} and H$_{2}$.  \ion{C}{4} emission is representative of the strength of the \ion{H}{1} Ly$\alpha$ line, which drives the observed flux level of H$_{2}$.  The non-variable nature of the lines leads to a Pearson correlation coefficient of 0.15, essentially uncorrelated.
 }
\end{center}
\end{figure}

\subsection{Young Brown Dwarfs: Low-Mass Classical T-Tauri Analogs}

As mentioned in the previous subsection, 2M1207 displays metal depletions consistent with those seen in some CTTSs.  The H$_{2}$ disk emission is also reminiscent of that observed around more massive young stars.   We therefore argue that 2M1207 is a low-mass analog to these systems.
While TW Hya is a somewhat atypical pre-main sequence object (with respect to the ages, accretion rates, and abundances of other CTTSs; we refer the reader to Section 1 of Herczeg et al. (2002) for a concise review), we use it for comparison with 2M1207 based on its well-studied far-UV spectrum~\citep{herczeg02}.  The H$_{2}$ emission seen in our COS observations is qualitatively similar to that of TW Hya, however there are quantitative differences in the far-UV spectra of these objects.  The first is the wealth of lines observed in the spectrum of TW Hya compared to 2M1207.  While the 2M1207 observations are at a lower S/N than the STIS observations of TW Hya,  there are numerous emission lines that would have been detected if they were present with the relative strengths seen in TW Hya (in particular, emission lines pumped by (0~--~2) R(0) 1217.21~\AA\ and  (0~--~2) R(1) 1217.64~\AA).  This implies that the Ly$\alpha$ emission profile in 2M1207 is considerably narrower than that observed in higher-mass CTTSs.
These ``missing'' fluorescent progressions are pumped by the wings of a broad stellar/shock Ly$\alpha$ emission profile, which are mostly inaccessible to COS due to contamination by geocoronal Ly$\alpha$.  The lack of a broad Ly$\alpha$ component in 2M1207 may be further evidence that Ly$\alpha$ is created in the accretion shock in this object~\citep{herczeg06}.  

We can make a quantitative comparison of the H$_{2}$ flux from TW Hya and 2M1207.  The total flux ratio between the two ($R^{TW}_{2M}$($TOT$)~$\equiv$~$I_{H2}$(TW Hya)/$I_{H2}$(2M1207)) is not the appropriate measure as TW Hya produces many more emission lines based on the broad stellar Ly$\alpha$ profile.  We compare the total H$_{2}$ emission from specific states observed in 2M1207, namely, those pumped by (1~--~2) R(6) 1215.73~\AA\ and (1~--~2) P(5) 1216.07~\AA.  The distance corrected flux ratios for the emission produced by pumping in those two lines are $R^{TW}_{2M}$(1~--~2 R(6))~=~391 and $R^{TW}_{2M}$(1~--~2 P(5))\footnote{Adding up the flux from the individual lines in TW Hya (Table 2 of Herczeg et al. 2002), we found a total flux of 369.6 $\times$~10$^{-15}$ ergs cm$^{-2}$ s$^{-1}$, in slight disagreement with the value of 350 quoted in their Table 6} =~350, respectively.  The $R^{TW}_{2M}$(1~--~2 R(6)) ratio is more susceptible to the effects of self-absorption by \ion{H}{1} in the circumstellar environment, though the ratios for both lines are similar.  
This implies that there is more Ly$\alpha$ flux per H$_{2}$ in the disk of TW Hya compared to the disk of 2M1207, assuming that the disk masses are proportional to the mass of the primary 
($M_{TW}$/$M_{2M}$~=~0.7 $M_{\odot}$/0.024 $M_{\odot}$~$\approx$~30).   The excess disk H$_{2}$ emission in TW Hya can be interpreted as a stronger local Ly$\alpha$ radiation field, which we propose is due to the higher mass accretion rate in TW Hya
($\sim$~2~$\times$ 10$^{-9}$ $M_{\odot}$ yr$^{-1}$ as compared to $\sim$~1~--~150~$\times$ 10$^{-12}$ $M_{\odot}$ yr$^{-1}$ for 2M1207; Herczeg et al. 2006, Scholz et al. 2005; Herczeg et al. 2009; this work) as well as a larger surface flux contribution from the magnetic, nonaccreting component on TW Hya.\nocite{herczeg06,scholz05,herczeg09} 

While these differences may reflect lower mass accretion rates in lower mass objects, the general trends connecting CTTSs and 2M1207 seem clear.  2M1207 is actively accreting from its disk, retains a warm (2500~--~4000 K) layer of H$_{2}$ in the inner disk, and shows evidence for depletion of Si and Mg into grains.  Given the edge-on geometry of the 2M1207 system, a more direct comparison would be to the edge-on CTTS DF Tau.  DF Tau was observed by COS as part of the $HST$ Cycle 17 Guaranteed Time program, and a comparison with 2M1207 will be presented in a future work.  If the additional H$_{2}$ component described in \S4.2 is attributable to an outflow, a better comparison might be made with the edge-on CTTS system DG Tau. \\ \\ \\ \\

\section{Summary}
We have presented far-UV spectroscopy of the young ($\sim$~10 Myr old) M8 brown dwarf / circumstellar disk system 2M1207.  These data provide an order of magnitude increase in spectral resolution over existing far-UV observations of a brown dwarf system.
We detect several emission lines of H$_{2}$ that are excited by Ly$\alpha$ photons created in an accretion shock, and use these lines to constrain the kinematics and physical state of the disk.  A second H$_{2}$ component exists, and we discuss possibilities for the origin of this emission, including at the stellar surface near the accretion shock and in an outflow.  A third possibility is that this H$_{2}$ feature is dayglow emission from the 6 $M_{J}$ giant planet, 2M1207b, however the data do not allow us to identify the exact location of the emitting region.
We measure several emission lines that trace the hot gas produced in the shock, including \ion{N}{5} and \ion{C}{4}.  Interestingly, we do not detect ions of refractory elements such as silicon and magnesium, and argue that these species have been depleted into grains.  This grain depletion scenario suggests that hot gas in the 2M1207 system is not significantly produced in a solar-type transition region, rather the accretion shock is responsible for the majority of the observed emission.  Although there are quantitative differences, these results suggest that young brown dwarfs harboring circumstellar disks are low-mass analogs to CTTSs.

\acknowledgments
It is a pleasure to acknowledge Greg Herczeg for valuable discussions at several phases of this project.  K.F. thanks Josh Destree and Charles Danforth for technical assistance with the COS LSF and Brian Keeney for reprocessing the FUV observations.  
This work made use of the CHIANTI atomic database, CHIANTI is a collaborative project involving the NRL (USA), the Universities of Florence (Italy) and Cambridge (UK), and George Mason University (USA).  This work was support by NASA grants NNX08AC146 and NAS5-98043 to the University of Colorado at Boulder.

\appendix

\section{\ion{O}{6} Flux Limits}

\ion{O}{6} can be an important shock temperature diagnostic, particularly when used in conjunction with the observed strengths of \ion{N}{5} and \ion{C}{4} (Danforth et al. 2001; Welsh et al. 2007 and references therein).~\nocite{danforth01,welsh07}  The production of these high ions in CTTSs is not fully understood~\citep{lamzin07}, hence in this appendix we use the COS observations to constrain the \ion{O}{6} emission produced in the 2M1207 system as a reference for future studies of low mass accreting systems.

During the Servicing Mission 4 Observatory Verification period, it was discovered that the MgF$_{2}$/Al mirrors of $HST$ have retained approximately 80 \% of their pre-flight reflectivity~\citep{mccandliss09}.  This (surprising) result has opened the door for use of the short wavelength response of the G140L mode of COS to perform spectroscopic observations at wavelengths inaccessible ($\lambda$ $<$ 1100~\AA) to previous $HST$ instruments~\citep{mccandliss09}.  The CALCOS pipeline processing of the G140L, segment B (400 $\lesssim$~$\lambda$~$\lesssim$~1150) is not yet mature enough to produce one-dimensional spectra appropriate for scientific analysis, however we performed a custom spectral extraction and reduction from the two-dimensional spectrograms that allowed us to create a low-resolution ($\Delta\lambda$~$\sim$~1.0~\AA) spectrum of 2M1207 from 912~--~1150~\AA.  We used this spectrum to set an upper limit on the integrated line strengths of the \ion{O}{6} $\lambda\lambda$ 1032, 1038~\AA\ resonance doublet (Table 3).    Assuming that any \ion{O}{6} produced in the accretion shock has the same velocity structure observed in the summed \ion{N}{5} and \ion{C}{4} profile, we find an 1-$\sigma$ upper limit to the \ion{O}{6} $\lambda$ 1032 and 1038~\AA\ emission to be 
5.1~$\times$ 10$^{-16}$ ergs cm$^{-2}$ s$^{-1}$.  

A more sophisticated approach to setting a limit on the level of \ion{O}{6} is to use the absence of \ion{O}{6} pumped H$_{2}$ emission lines in the higher sensitivity M-grating observations.  We can set limits on the amount of \ion{O}{6} emission that may be emitted from the accretion shock by making rough assumptions about the total column density of the emitting H$_{2}$, $N$(H$_{2}$).  This method follows the analysis of \ion{O}{6} pumped H$_{2}$ in the circumstellar disk around the M1V star AU Microscopii presented by~\citet{france07b}.  The first step is to identify the COS band H$_{2}$ lines that offer the most stringent upper limits, which will be a combination of intrinsic line strength and COS sensitivity at the corresponding wavelength.   For this purpose, the tightest limit is set by the non-detection of the $C$~--~$X$ (1~--~4) Q(3) H$_{2}$ line at 1163.81~\AA.  
Using the measured upper limit on this \ion{O}{6} pumped H$_{2}$ emission line flux (1.1 $\times$ 10$^{-6}$ photons s$^{-1}$ cm$^{-2}$), we can calculate a limit to the total fluorescent output from the \ion{O}{6} pumped cascade.  The total emitted flux out of the electrovibrational state ($n',v',J'$),  $\sum\limits_{j} F_{ij}$, is given by
\begin{equation}
\sum\limits_{j} F_{ij}~=~F_{ik}~\left(~\frac{ A_{ik} }{ \sum\limits_{l} A_{il}}~\right)^{-1}~(1-\xi_{i})^{-1} 
\end{equation}
where $i$ refers to the upper state ($n'$,$v'$,$J'$).  The indices $j$, $k$, and $l$ refer to the lower states ($n''$,$v''$,$J''$).  
$F_{ik}$ is the upper limit of the flux in the COS band lines (in photons s$^{-1}$ cm$^{-2}$).
The ratios of individual to total Einstein $A$-values~\citep{abgrall93b} are the branching ratios, and $\xi_{i}$ is a correction for the efficiency of predissociation in the excited electronic state~\citep{liu96}.  
In the case of emission excited by coincidence with \ion{O}{6} , we are concerned with Werner band emission 
($n'$~--~$n''$ = $C$~--~$X$~$\equiv$~$C^{1}\Pi_{u}$~--~$X^{1}\Sigma^{+}_{g}$), $v'$ = 1, and $v''$ = 4  for the 1163.81 \AA\ line.   In the general form, Equation A1 must also be summed over the possible 
redistribution over rotational ($\Delta J$ = $\pm$1, 0) states, however, the present case
is simplified due to the parity selection rules that forbid Q branch 
($\Delta J$ = 0) transitions to mix with R ($\Delta J$ = -1) and P ($\Delta J$ = +1) branches.  
The predissociation fraction for the Werner bands is zero ($\xi_{C}$ = 0; Ajello et al., 1984).~\nocite{ajello84}

Following this procedure, we arrived at the total emitted photon flux, 
derived from the observed (1~--~4) 1163.81 \AA\ upper limit.  
Applying Equation A1, we find that the limit to the total emitted flux from the \ion{O}{6} pumped H$_{2}$ cascade is $\sum\limits_{j} F_{ij}$ $\leq$ 4.0 $\times$ 10$^{-6}$ photons s$^{-1}$ cm$^{-2}$.  
In order to convert this emission upper limit into a flux limit on the exciting \ion{O}{6} $\lambda$~1032~\AA\ line, some assumptions must be made regarding the geometry and characteristics of the H$_{2}$ population.  The H$_{2}$ covering fraction will likely be 0.5 or less, depending on the standoff distance between the \ion{O}{6} emitting region and the inner edge of the disk.  We will use 0.5 for the purposes of this calculation.
In Section 3.1, we show that the temperature of the molecular phase of the circumstellar disk has 2500~$\lesssim$~$T$(H$_{2}$)~$\lesssim$~4000 K, so we assume a thermal width of 2500 K for the H$_{2}$ absorption at \ion{O}{6}.  The upper limit to the exciting flux is found by balancing the number of absorbed photons with the maximum line flux that can be accommodated without producing a detectable level of H$_{2}$ emission.  A total column density of $N$(H$_{2}$)~$\approx$~ 3~$\times$~10$^{16}$ cm$^{-2}$ is a rough estimate for the 2M1207 circumstellar disk value.  This is the value where the [$v$,$J$] = [1,3] absorption line begins to saturate (for a 2500 K population), meaning that the total number of absorbed photons rises slowly from this column until damping wings become present, at the unrealistically large total column of $\sim$~10$^{20}$ cm$^{-2}$.  
This total H$_{2}$ column corresponds to a column density in the $N$($v$,$J$) = $N$(1,3) state of 5.7~$\times$~10$^{14}$ cm$^{-2}$.  Using the peak of the \ion{O}{6} emission as a free parameter, we find that the total number of photons absorbed reaches the upper limit on the total number of emitted photons for an \ion{O}{6} $\lambda$~1032~\AA\ line strength 
of $I_{1032}$ $<$ 9.5~$\times$~10$^{-16}$ ergs cm$^{-2}$ s$^{-1}$.   We see that given the number of assumptions made, one cannot place a more stringent limit on the \ion{O}{6} flux than using the G140L segment B observation.  If one had an independent measure of the column density (from near- or mid-IR rovibrational emission lines for example), this method would be far more robust, however it is very unlikely that these lines could be detected with current IR instruments (e.g. Lupu et al. 2006; France et al. 2007).~\nocite{france07b,lupu06}


\bibliography{ms}




\end{document}